\documentclass[8.5pt,twoside,twocolumn]{article}
\usepackage[super,sort&compress,comma]{natbib}
\usepackage{mhchem}
\usepackage{times,mathptmx}
\usepackage{sectsty}
\usepackage{balance}
\usepackage{color}
\usepackage{graphicx} %eps figures can be used instead
\usepackage{lastpage}
\usepackage[format=plain,justification=raggedright,singlelinecheck=false,font=small,labelfont=bf,labelsep=space]{caption}
\usepackage{fancyhdr}
\begin{document}

\thispagestyle{plain}
\fancypagestyle{plain}{
%\fancyhead[L]{\includegraphics[height=8pt]{headers/LH}}
%\fancyhead[C]{\hspace{-1cm}\includegraphics[height=20pt]{headers/CH}}
%\fancyhead[R]{\includegraphics[height=10pt]{headers/RH}\vspace{-0.2cm}}
\renewcommand{\headrulewidth}{1pt}}
\renewcommand{\thefootnote}{\fnsymbol{footnote}}
\renewcommand\footnoterule{\vspace*{1pt}%
\hrule width 3.4in height 0.4pt \vspace*{5pt}}
\setcounter{secnumdepth}{5}

\makeatletter
\def\subsubsection{\@startsection{subsubsection}{3}{10pt}{-1.25ex plus -1ex minus -.1ex}{0ex plus 0ex}{\normalsize\bf}}
\def\paragraph{\@startsection{paragraph}{4}{10pt}{-1.25ex plus -1ex minus -.1ex}{0ex plus 0ex}{\normalsize\textit}}
\renewcommand\@biblabel[1]{#1}
\renewcommand\@makefntext[1]%
{\noindent\makebox[0pt][r]{\@thefnmark\,}#1}
\makeatother
\renewcommand{\figurename}{\small{Fig.}~}
\sectionfont{\large}
\subsectionfont{\normalsize}

\fancyfoot{}
%\fancyfoot[LO,RE]{\vspace{-7pt}\includegraphics[height=9pt]{headers/LF}}
%\fancyfoot[CO]{\vspace{-7.2pt}\hspace{12.2cm}\includegraphics{headers/RF}}
%\fancyfoot[CE]{\vspace{-7.5pt}\hspace{-13.5cm}\includegraphics{headers/RF}}
%\fancyfoot[RO]{\footnotesize{\sffamily{1--\pageref{LastPage} ~\textbar\hspace{1pt}\thepage}}}
\fancyfoot[RO]{\footnotesize{\sffamily{1--\pageref{LastPage} ~\textbar\hspace{1pt}\thepage}}}
\fancyfoot[LE]{\footnotesize{\sffamily{\thepage~\textbar\hspace{0cm} 1--\pageref{LastPage}}}}
\fancyhead{}
\renewcommand{\headrulewidth}{1pt}
\renewcommand{\footrulewidth}{1pt}
\setlength{\arrayrulewidth}{1pt}
\setlength{\columnsep}{6.5mm}
\setlength\bibsep{1pt}

\twocolumn[
  \begin{@twocolumnfalse}
\noindent\LARGE{\textbf{Tunable Half-metallic Magnetism in Atom-thin Holey Two-dimensional C$_2$N Monolayer}}
\vspace{0.6cm}

\noindent\large{\textbf{Sai Gong,\textit{$^{a}$} Wenhui Wan,\textit{$^{a}$} Shan Guan,\textit{$^{a,b}$} Bo Tai,\textit{$^{b}$} Chang Liu,\textit{$^{a}$} Botao Fu,\textit{$^{a}$} Shengyuan A. Yang,\textit{$^{\ast b}$}and Yugui Yao\textit{$^{\ast a}$}}}\vspace{0.5cm}
%Please note that \ast indicates the corresponding author(s) but no footnote text is required.

%\noindent\textit{\small{\textbf{}}}

%\noindent \textbf{\small{}}
%\vspace{0.6cm}
%Please do not change this text.

\noindent \normalsize{Exploring two-dimensional (2D) materials with magnetic ordering is a focus of current research. It remains a challenge to achieve tunable magnetism in a material of one-atom-thickness without introducing extrinsic magnetic atoms or defects. Here, based on first-principles calculations, we propose that tunable ferromagnetism can be realized in the recently synthesized holey 2D C$_2$N ($h$2D-C$_2$N) monolayer via purely electron doping that can be readily achieved by gating. We show that owing to the prominent van Hove singularity in the band structure, the material exhibits spontaneous ferromagnetism at a relatively low doping density. Remarkably, over a wide doping range of 4$\times$10$^{13}$/cm$^2$ to 8$\times$10$^{13}$/cm$^2$, the system becomes half-metallic, with carriers fully spin-polarized. The estimated Curie temperature can be up to 320 K. Besides gating, we find that the magnetism can also be effectively tuned by lattice strain. Our result identifies $h$2D-C$_2$N as the first material with single-atom-thickness that can host gate-tunable room-temperature half-metallic magnetism, suggesting it as a promising platform to explore nanoscale magnetism and flexible spintronic devices.}
\vspace{0.5cm}
 \end{@twocolumnfalse}
  ]

\section{Introduction}
Two-dimensional (2D) materials have been attracting tremendous interest since the discovery of graphene in 2004.\cite{1} Many new 2D materials have been realized to date.\cite{4,5,6} In this family of 2D materials, graphene still stands out in that it is completely flat with only one-atom-thickness. Other monolayer materials such as silicene or phosphorene have buckled structures with atoms displaced perpendicular to the 2D plane; while materials like transition metal dichalcogenides consist of multiple atomic layers. The single-atom-thickness endows graphene with superior properties.\cite{3} Representing the ultimate limit of a material's thickness, it enables devices that are more compact in size; and the atomic thickness also makes the material more susceptible to external control, e.g., the carrier density in graphene can be tuned up to $10^{14}$/cm$^2$ by gating.\cite{57,58}

Meanwhile, it is much desired to realize various symmetry-breaking phases in 2D materials. 2D materials with magnetic ordering are of particular interest, because such nanoscale magnetism is not only fascinating from a scientific point of view, it is also of great value for important applications such as magnetic information storage and spintronics devices.\cite{wolf,hanw} Extensive efforts have been devoted in this direction, mostly by doping non-magnetic 2D materials with magnetic atoms or by creating certain structural defects.\cite{9,h2,h3,caoc,yccheng,yjia} However, the induced magnetism is usually weak, and it is very difficult to control the distribution of dopants or defects in experiment. Moreover, acting as extrinsically introduced scattering centers, these dopants or defects would strongly scatter the carriers and deteriorate the material's transport properties.

It would become much better if we could generate magnetism without introducing extrinsic dopants or defects, which could avoid the above-mentioned severe disadvantages. It was predicted that the 2D MoN$_2$ is an intrinsic magnetic material with its magnetic ordering tunable by strain.\cite{n4,sswang} But monolayer MoN$_2$ has not been experimentally realized yet. Another interesting proposal is to achieve tunable ferromagnetism in monolayer GaSe via gate controlled hole-doping.\cite{38,wyao} The resulting spin-polarization can be large, however, the magnetic transition (Curie) temperature is still low ($\sim 90$ K), not high enough for practical applications. It hence remains a challenge to find a dopant-free 2D material with strong magnetism. In addition, these magnetic 2D crystals consist of multiple atomic layers, not being atomically thick. One may further ask: Can we push the 2D magnetism down to one-atom-thickness?

In this work, we address the above question by proposing half-metallic magnetism in holey 2D C$_2$N ($h$2D-C$_2$N) monolayer crystal with single-atom-thickness via purely electron doping that can be readily achieved by gating. $h$2D-C$_2$N is a newly synthesized 2D material. Experimentally, monolayer $h$2D-C$_2$N has been successfully isolated and transferred onto insulating substrate.\cite{40} It shows a direct bandgap $\sim$1.96 eV measured by optical spectroscopy, and a field-effect-transistor fabricated based on it exhibits a high on/off ratio of $10^7$.\cite{40} Without charge doping, $h$2D-C$_2$N is non-magnetic. Here, using first-principles calculations, we show that ferromagnetic ordering can be easily induced in $h$2D-C$_2$N monolayer by electron doping, and remarkably, in a wide doping range ($4\times 10^{13}$/cm$^2$ to $8\times 10^{13}$/cm$^2$) the material is in fact a half-metal with fully spin-polarized carriers for conduction. Such doping levels can be achieved with currently available gating techniques, and gating provides a highly-efficient method to control magnetism in this material. Furthermore, we find that the magnetism can also be effectively tuned by lattice strain: starting from a half-metal state, a uniaxial strain $\sim 10\%$ can completely turn off the magnetism. This is the first time that half-metallic ferromagnetism is ever proposed in a one-atom-thin 2D material via an intrinsic approach, without magnetic dopants or defects. These important advantages---the atomic thickness, the half metallicity, the easy control via gating or strain, and the experimental availability of the material---will make $h$2D-C$_2$N a highly promising platform to explore nanoscale magnetism and the many electronic and spintronic device applications.

\begin{figure*}[htb!]
\centerline{\includegraphics[width=0.8\textwidth]{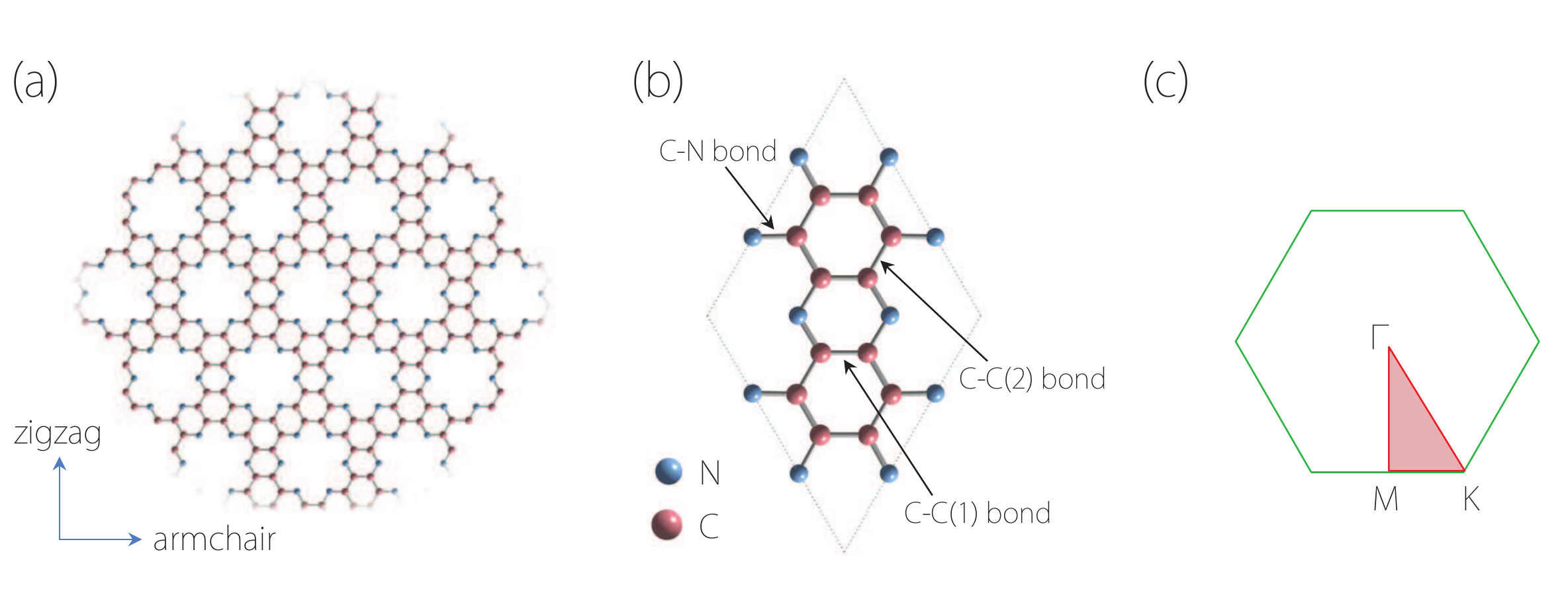}}
\caption{(a) Top view of the lattice structure of a $h$2D-C$_2$N monolayer. The structure is flat with single-atom-thickness. (b) Primitive cell of $h$2D-C$_2$N monolayer, in which the three different bond lengths are labeled. (c) Schematic picture of the first Brillouin zone.  }\label{fig1}
\end{figure*}

\section{Computational methods}
Our first-principles calculations are based on the density functional theory (DFT) using the projector augmented wave (PAW) method\cite{44} as implemented in the Vienna $ab$-$initio$ simulation package (VASP).\cite{45,46} The generalized gradient approximation with the Perdew-Burke-Ernzerhof (PBE) realization\cite{pbe} is used for the exchange-correlation functional. Our main results are further checked by the hybrid functional (HSE06)\cite{hse} approach. The kinetic energy cutoff is set to be 550 eV. The lattice constants and the ionic positions are fully optimized until the force on each ion is less than 0.01 eV/\AA,  and the structural optimization is performed with a smearing of 0.05 eV. The convergence criteria for energy is chosen as 10$^{-5}$ eV. And a vacuum layer of 18 \AA\ thickness is added to avoid artificial interactions between periodic images. The effect of charge-doping is simulated by changing the total number of electrons in the unit cell, with a compensating jellium background of opposite charge to maintain the neutrality. In calculating magnetic properties, we use a refined $k$-mesh of size 27$\times$27$\times$1 to ensure the convergence of result,  and the tetrahedron method with Bl\"{o}chl corrections is adopted for the Brillouin zone integration.

\section{Results and discussion}
\subsection{Electronic structure}

The lattice structure of monolayer $h$2D-C$_2$N is shown in figure~\ref{fig1}, which may be viewed as a honeycomb lattice of benzene rings connected through nitrogen atoms. The lattice is completely flat, with only one atom thickness.
The primitive unit cell consists of 12 C and 6 N atoms (see figure~\ref{fig1}(b)), making a C$_2$N stoichiometry. The optimized lattice parameter that we obtain is 8.330 \AA, which agrees well with the experiment measurement and
previous theoretical studies.\cite{40,41,42,lzhu,43} The structure involves three types of bonds, as indicated in figure~\ref{fig1}(b). We find that C-N bond $\approx1.339$ \AA, C-C(1) bond $\approx1.432$ \AA, and C-C(2) bond $\approx1.469$ \AA.

\begin{figure*}[htb!]
\centerline{\includegraphics[width=0.8\textwidth]{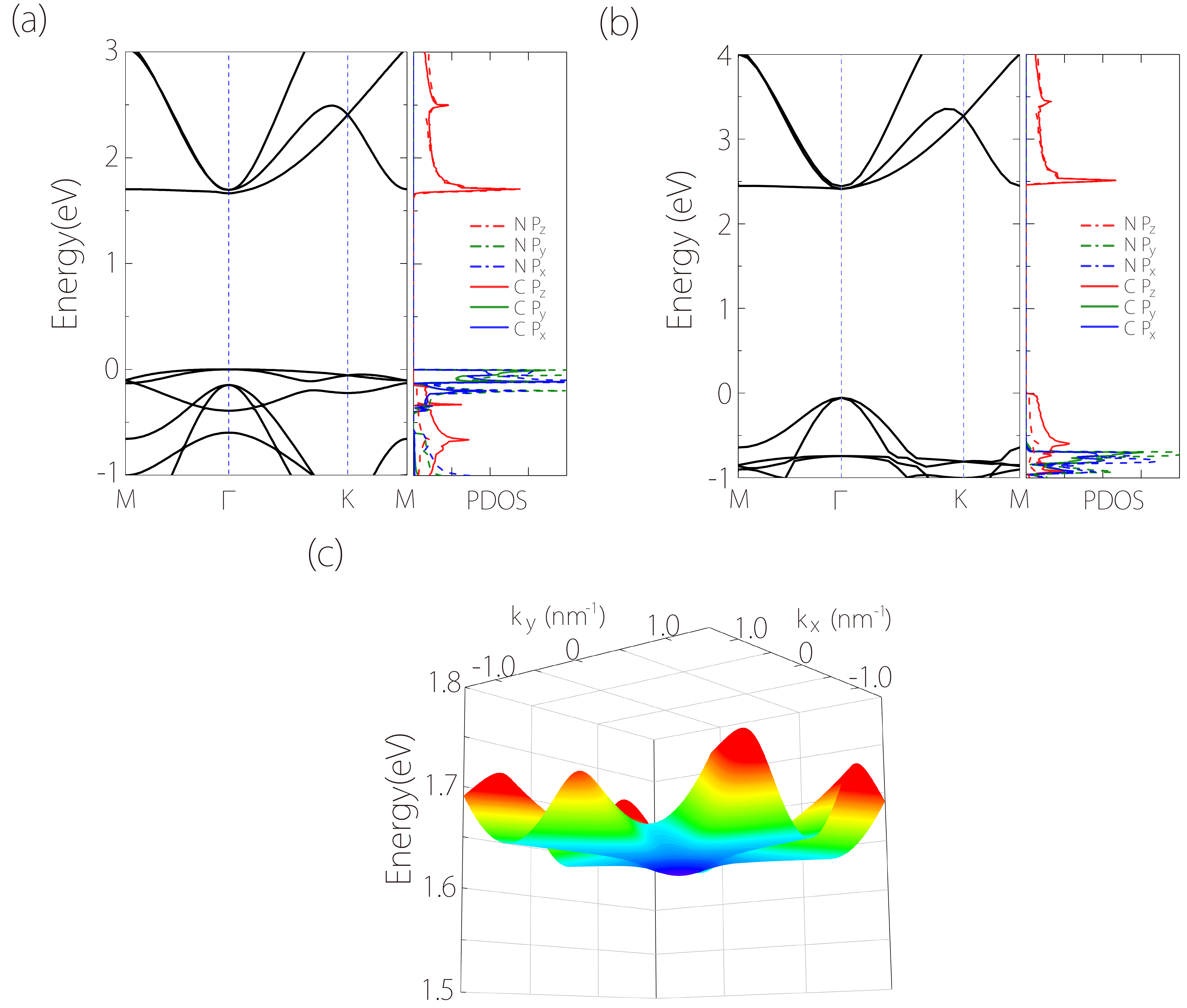}}
\caption{(a,b) Band structure of $h$2D-C$_2$N at zero doping along with the orbital projected density of states (DOS) obtained using (a) PBE approach,  and (b) HSE06 approach. (c) 3D plot of the band dispersion near the conduction band edge, exhibiting almost flat dispersion along the $\Gamma$-M directions which contributes to the pronounced van Hove singularity as shown in (a). }\label{fig2}
\end{figure*}

The calculated electronic band structure with zero doping is shown in figure~\ref{fig2}(a).  It shows a semiconducting phase with a direct bandgap $\approx1.65$ eV at the $\Gamma$ point, in good agreement with the previous DFT results and the experimental value ($\sim1.96$ eV).\cite{40,41,43} The conduction band minimum (CBM) at the $\Gamma$ point involves three nearly degenerate states, and one band has almost flat dispersion along the $\Gamma$-M direction (see figure~\ref{fig2}(c)). This leads to a prominent van Hove singularity in the density of states (DOS) at the CBM. In the right panel of figure~\ref{fig2}(a), we show the projected DOS. One observes that the states contributing to the van Hove singularity at CBM are mainly from the $p_z$ orbitals of C and N atoms, i.e., from the $\pi$ electrons.  Compared with graphene, which has massless linear dispersion for the $\pi$ electrons, the appearance of almost flat dispersion which contributes to the singular DOS in $h$2D-C$_2$N should have a strong connection with the holey structure. Such van Hove singularity and the associated huge DOS would generally lead to instabilities towards symmetry-breaking phase transitions, which is our focus in this work.

We mention that nearly flat bands and van Hove singularity also appear at the valence band maximum (VBM) in the PBE band structure, where the states are mainly consisting of $p_x$ and $p_y$ orbitals. A recent work studied the possible hole-doping-induced magnetism associated with this VBM van Hove singularity.\cite{Liang2016} However, with more accurate hybrid functional (HSE06) approach, it is found that the band ordering near VBM is completely changed from the PBE result, as shown in figure~\ref{fig2}(b). Particularly, the van Hove singularity is pushed below the VBM on HSE level, therefore suppressing the hole-doping-induced magnetism. Meanwhile, the HSE band structure is almost the same as the PBE result around CBM. Thus, in this work, we will focus on the CBM and consider the case of electron-doping. Our subsequent presentation will be based on the PBE result for the conduction bands (main results are also confirmed by HSE calculations).

\subsection{Electron doping}
\begin{figure*}[!htb]
\centerline{\includegraphics[width=0.8\textwidth]{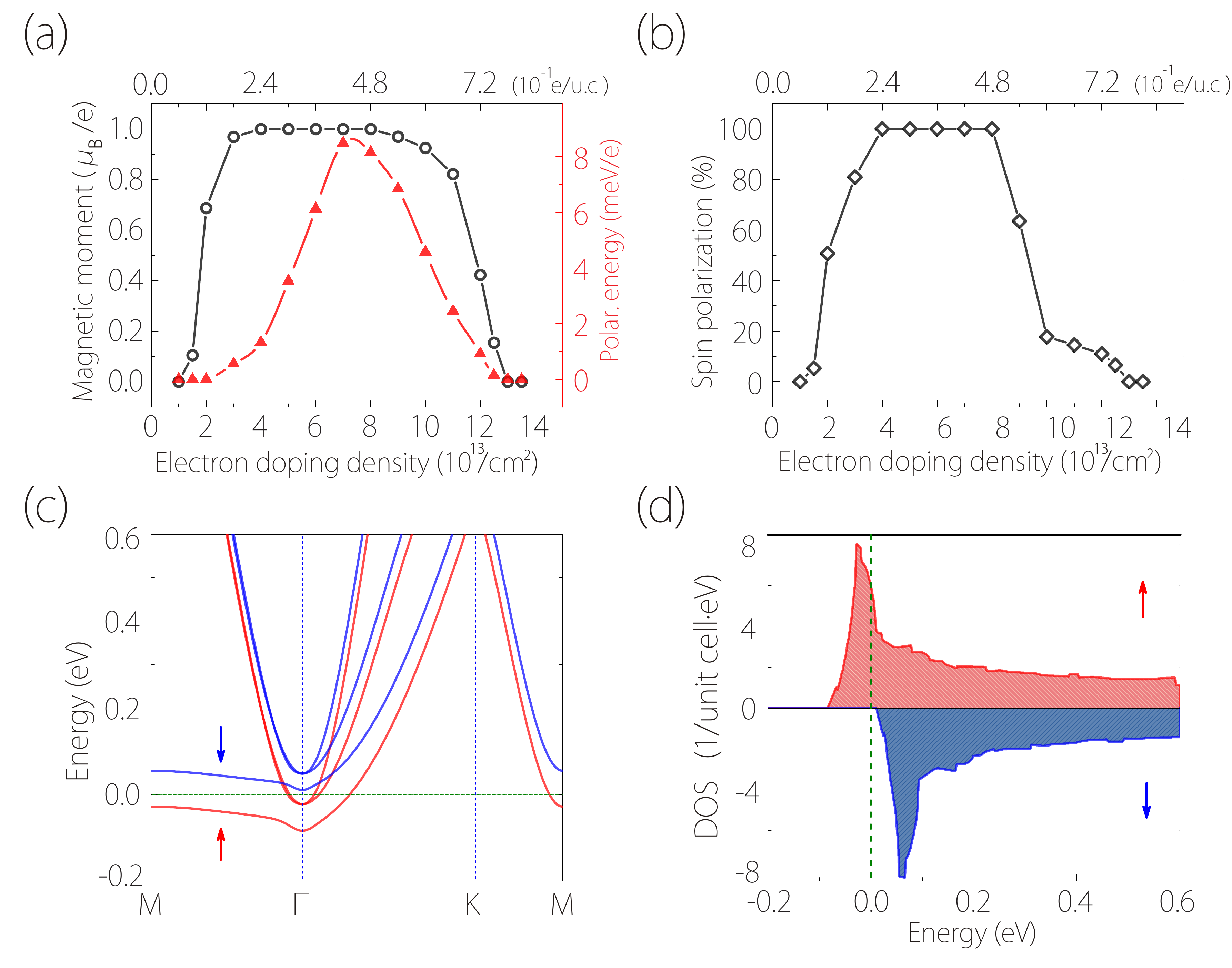}}
\caption{(a) Magnetic moment (black circles)
and spin polarization energy (red triangles) per electron carrier versus the doping density.
(b) Spin polarization of the carriers as a function of doping density.
(c) Spin-resolved band structure (around the Fermi level) at the electron doping level of 6$\times$10$^{13}$/cm$^2$.  (d) Spin-resolved density of states (DOS) corresponding to the band structure in (c). Here the Fermi level is taken to be energy zero.  }\label{fig3}
\end{figure*}

Large DOS associated with van Hove singularities is usually connected with instabilities towards symmetry-breaking phases. At zero doping ($n=0$, where $n$ is the electron doping density), the $h$2D-C$_2$N monolayer is in the non-magnetic state. However, due to the van Hove singularity at CBM, we find that the system develops a spontaneous ferromagnetic ordering even with a slight electron doping.  (In calculation, we also checked the possibility of antiferromagnetic ordering, which turns out to be energetically higher than the ferromagnetic state, as shown in the Supporting Information.) In figure~\ref{fig3}(a), we plot the magnetic moment and the spin polarization energy (both are per doped electron carrier) versus the doping density. Here the spin polarization energy is defined as the energy difference between the non-spin-polarized phase and the spin-polarized phase. One observes that ferromagnetism already appears at a doping density of $n=1.5\times 10^{13}$/cm$^2$  (here $1\times 10^{13}$/cm$^2\approx 0.06$e$/$unit cell (u.c.)), and with increasing doping density, the magnetic moment per carrier has a sharp increase, reaching 0.96$\mu_B$/e at $n=3\times 10^{13}$/cm$^2$. The moment is close to 1$\mu_B$ per carrier in a wide range from $n=4\times 10^{13}$/cm$^2$ to $8\times 10^{13}$/cm$^2$, indicating that the carriers become fully spin-polarized in this doping range. Above the range, the moment begins to decrease and has a quick drop above $n=1\times 10^{14}$/cm$^2$. In comparison, the spin polarization energy also increase first and then decrease as a function of $n$, however, it does not show a plateau region like the magnetic moment. It has a peak around $n=7\times 10^{13}$/cm$^2$ with a value of $\approx 8.5$ meV per electron carrier. This value is quite large, almost three times of that in monolayer GaSe,\cite{38} suggesting that the ferromagnetic ordering here would be more stable. The spin polarization energy drops to zero above $1.3\times 10^{14}$/cm$^2$, indicating that the system returns to non-magnetic at higher doping.

In figure~\ref{fig3}(b), we plot the spin polarization of the carriers $P=(n_\uparrow-n_\downarrow)/(n_\uparrow+n_\downarrow)$, where $n_\uparrow$ ($n_\downarrow$) is the population of spin-up (spin-down) carriers.  Here the polarization is defined for the occupied conduction band states at zero temperature. One clearly observes that the carriers are 100\% spin-polarized in the doping range from $n=4\times 10^{13}$/cm$^2$ to $8\times 10^{13}$/cm$^2$, where the system becomes half-metallic. As a representative, we take $n=6\times 10^{13}$/cm$^2$ in this range, and plot the corresponding band structure in figure~\ref{fig3}(c) and the spin-resolved DOS in figure~\ref{fig3}(d). Compared with the undoped (non-magnetic) case, one observes that the shape of the band dispersion is more or less unchanged. The main difference is that the bands are now spin-split, and the Fermi level cuts through only one spin species. The energy splitting between the two spin bands around CBM is about 0.1 eV. Such half-metal state with 100\% spin polarization allows fully spin-polarized transport, hence it is very much desired for spintronics applications. Our result shows that half-metallic ferromagnetism can indeed be realized in a material with only one atom thickness without extrinsic dopant atoms or defects.

The emergence of ferromagnetism here can be explained in the Stoner picture,\cite{54} which has been adopted to explain the magnetism in several $p$-orbital systems, particularly in so-called ``$d^0$ ferromagnetic semiconductors''.\cite{hpan,56} In this model, ferromagnetism would appear spontaneously if the Stoner criterion $IN(E_F)>1$ is satisfied. Here $I$ is the strength of the magnetic exchange interaction, and $N(E_F)$ is the DOS at the Fermi level of the non-spin-polarized band structure. When the Stoner parameter $IN(E_F)$ is greater than 1, the relative gain in exchange energy would dominate over the loss in kinetic energy, and hence the system would favor a ferromagnetic ground state. Indeed, in our DFT calculation, we find that the Stoner parameter is larger than 1 in a wide range from $\sim 1.5\times 10^{13}$/cm$^2$ up to $\sim 1.25\times 10^{14}$/cm$^2$ (its value is about 4 in the half-metallic range). This analysis indicates that the huge DOS associated with the van Hove singularity plays a key role in the appearance of ferromagnetism.

 It is important to note that besides magnetic instability, a large DOS may also lead to instability towards structural distortions. We have investigated such possibility by relaxing the structure (with finite doping) from slightly randomized atomic positions, and verified that the relaxed structure has little difference from the undoped one. Furthermore, we have calculated the phonon spectrum at finite doping. The result has no imaginary frequency (see Supporting Information), showing that the structure under doping remains stable.

As we mentioned, the electron doping level can be conveniently tuned by the available gating technique. For example, the doping density on the order of $10^{14}$/cm$^2$ has been demonstrated in graphene by ion liquid gating,\cite{57,58} and of $10^{13}$/cm$^2$ in transition metal dichalcogenides by back-gate gating.\cite{mak,yzhang} Owing to its atomic-thickness, we expect that doping range discussed in previous paragraphs can be readily achieved in $h$2D-C$_2$N monolayers. We have also estimated the magnetic transition (Curie) temperature. At the mean field level, this can be done by evaluating the magnetic moment through minimizing the electronic free energy at a given temperature~\cite{38}. We find that the estimated Curie temperature can be up to 320 K at the doping density of $8\times 10^{13}$/cm$^2$, correlated with the large spin-polarization energy discussed in figure~\ref{fig3}(a). One should be cautious that the mean field approach in general overestimates the transition temperature. For low-dimensional systems, the fluctuation effects usually lead to a strong reduction of the Curie temperature. Nevertheless, the results here at least indicate that a strong half-metallic ferromagnetism can be achieved in an existing atom-thin 2D material, and this magnetism can be readily turned on and off via gating.

\begin{figure}[htb]
\centerline{\includegraphics[width=0.5\textwidth]{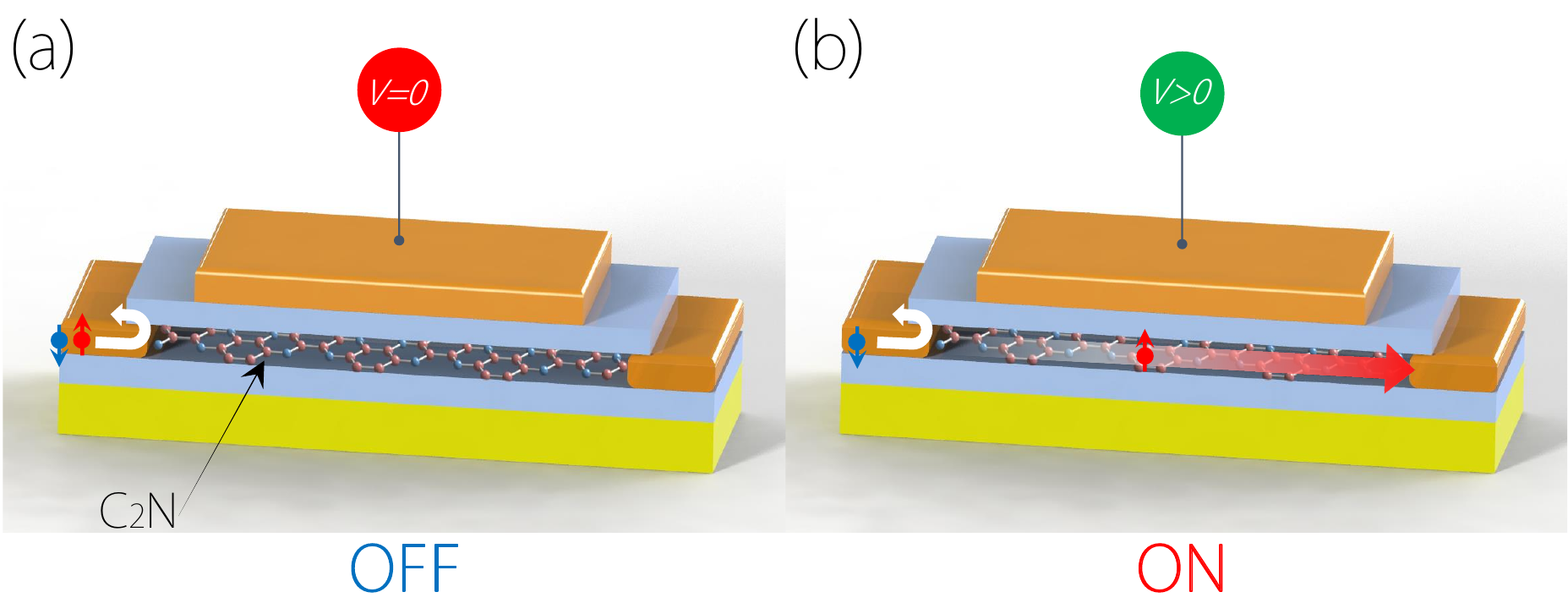}}
\caption{ Schematic of a spin-polarized field-effect transistor device using C$_2$N monolayer as the channel material. (a) Without gate voltage, the  channel is insulating. No current can pass through. (b) By applying a gate voltage, the C$_2$N monolayer is driven to the half-metallic phase. Hence a fully spin-polarized current can flow through.}\label{fig4}
\end{figure}

An experimental setup that can be used to probe the above predicted property is illustrated in figure~\ref{fig4}. It takes a standard field effect transistor configuration, with a $h$2D-C$_2$N monolayer as the channel material. Without applied gate voltage (figure~\ref{fig4}(a)), the channel is a good semiconductor, and no current can pass through. Hence the transistor is in the off state. When a gate voltage is applied (figure~\ref{fig4}(b)), the $h$2D-C$_2$N monolayer becomes half-metallic. Then current can flow through, and most importantly, the current is fully spin-polarized. The spin-polarization can be measured by using magnetic electrode or by subsequent spin filtering. In fact, the device proposed here can be well incorporated into spintronics circuits as a controlled spin generator. It has the advantages of compactness, small operating voltage, and low energy consumption.

\subsection{Strain effects}
\begin{figure*}[htb]
\centerline{\includegraphics[width=0.8\textwidth]{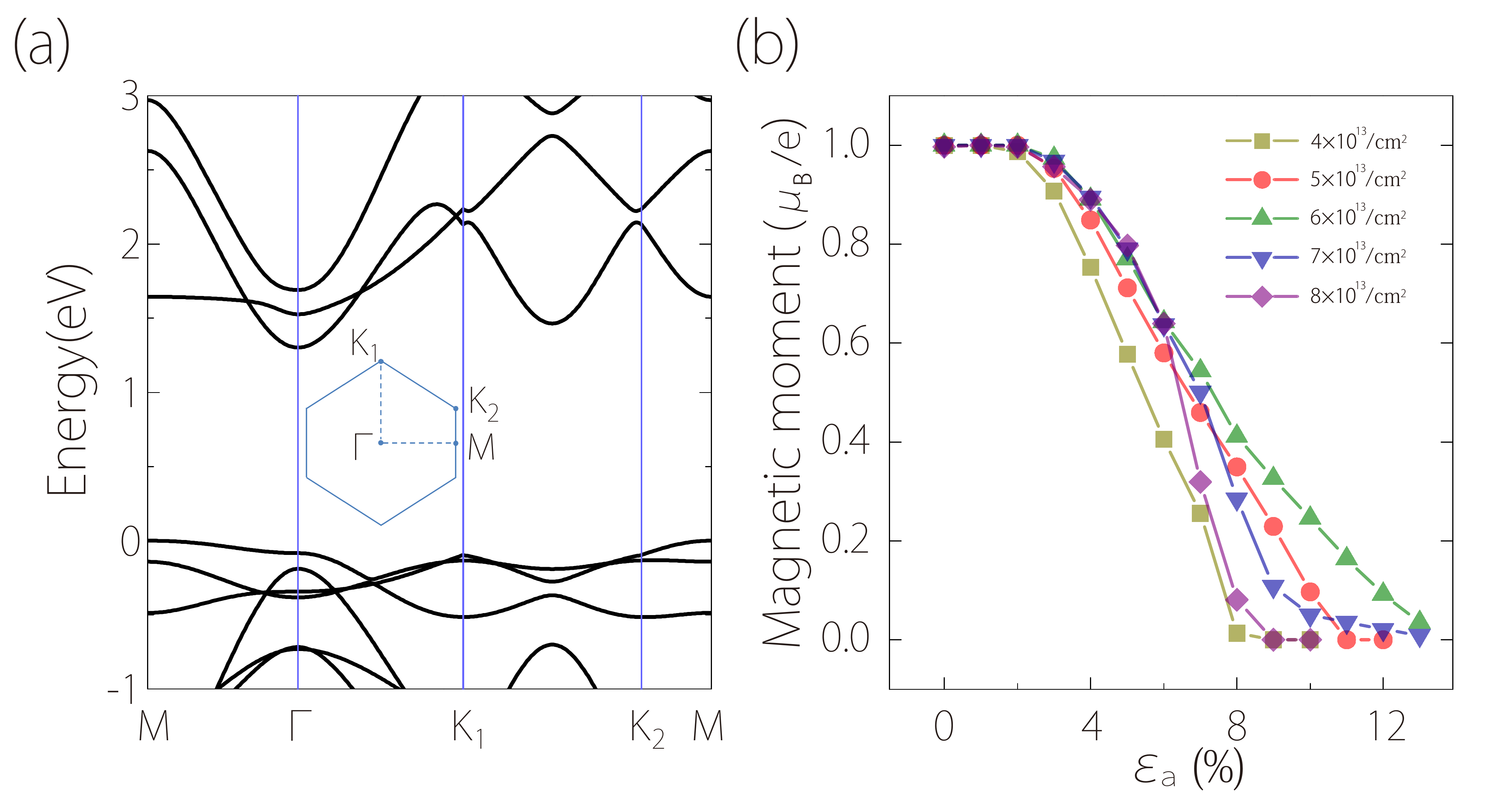}}
\caption{(a) Band structure of $h$2D-C$_2$N monolayer at zero doping under a uniaxial strain $\varepsilon_a$ of $8\%$. (b) Magnetic moment per carrier versus strain $\varepsilon_a$ for several different doping densities. Here the strain $\varepsilon_a$ is applied along the armchair direction.}\label{fig5}
\end{figure*}

Besides gating, the magnetic property may also be tuned by lattice strain. 2D materials typically can sustain large strains, e.g., graphene, MoS$_2$, and phosphorene can sustain strains above $25\%$.\cite{strain1,strain2,strain3,strain4} A previous computational study has shown that the critical strain of $h$2D-C$_2$N monolayer is larger than $12\%$.\cite{43} In figure~\ref{fig5}(a), we plot the band structure of $h$2D-C$_2$N monolayer without doping under a uniaxial 8\% strain along the armchair direction. Compared with the unstrained band structure in figure~\ref{fig2}(a), one observes that the initially degenerate states at the CBM split due to the reduced symmetry, and a more dispersed band is pushed below the flat band, forming the new CBM. As a result, the original van Hove singularity is higher in energy and the DOS at the new CBM is much decreased. According to the Stoner model, this would reduce the tendency towards ferromagnetism under electron doping. Indeed, as shown in figure~\ref{fig5}(b), starting from an initially half-metallic state, e.g., with $n=4\times 10^{13}$/cm$^2$, by increasing uniaxial strain, the magnetic moment per carrier is decreased continuously, and is fully suppressed above 8\% strain. Similar trend is observed at other doping densities and also for strains along the zigzag direction (see Supporting Information). Experimentally, strain can be applied by stretching or bending a flexible substrate on which the 2D material is attached, or it can be controlled electrically by using a piezoelectric substrate.\cite{akin,chang,yhui} It provides a powerful alternative method to control the magnetism in $h$2D-C$_2$N monolayers.

\subsection{Discussion}
We discuss several points before concluding. First, in the above calculation, we take the spin-polarization direction to be perpendicular to the 2D plane. Due to the very weak spin-orbit coupling (SOC) strength of the light elements C and N, the material exhibits very weak magneto-crystalline anisotropy. We have performed the calculation including SOC and found that the out-of-plane polarized state and the in-plane polarized state have negligible energy difference (less than $0.02$ meV/e). This shows that the material is a soft ferromagnet, with a low magnetic coercivity. This could be an important advantage for its use as dynamic layers in nano-magnetic devices, which leads to low energy dissipation during magnetic reversal processes.

Second, a recent theoretical study proposed to realize magnetism in $h$2D-C$_2$N by adsorption of magnetic atoms.\cite{yjia} From the analysis here, we can see that not only the localized magnetic moments of the adsorbed magnetic atoms contribute to the magnetism, the possible charge doping could also contribute. In fact, the former factor is not a necessity for the induced magnetism in the current system. We have checked that even by adsorption of Alkali atoms, due to the charge doping, the $h$2D-C$_2$N monolayer still becomes magnetic (see Supporting Information).
In addition, as we mentioned in the introduction part, introducing extrinsic dopant atoms has several severe drawbacks. The gate-controlled charge doping proposed here would be a more preferred approach.

Finally, when $h$2D-C$_2$N monolayers are stacked along the perpendicular direction, they form multilayer structures bonded by weak van der Waals interactions.\cite{40} Hence one expects that the doping induced magnetism may also appear for $h$2D-C$_2$N multilayers. It is also possible to combine $h$2D-C$_2$N with other 2D material into van der Waals heterostructures,\cite{vdwh} in which the interaction between magnetism and other possible orderings and carrier dynamics could lead to intriguing effects and device functionalities.
	
\section{Conclusion}
In conclusion, by using first-principles calculations, we show that tunable ferromagnetism and half-metallicity can be generated in the recently synthesized C$_2$N-$h$2D monolayers via charge doping controlled by electric gating. The half-metallic state can be realized in a wide range of doping density from 4$\times$10$^{13}$/cm$^2$ to 8$\times$10$^{13}$/cm$^2$, achievable by the existing gating technique. The Curie temperature is estimated up to 320 K at mean field level. In addition, the magnetism can be effectively tuned using lattice strain. This is the first time that half-metallic magnetism is ever proposed in a 2D material with one-atom-thickness, which opens great opportunity to explore nanoscale magnetism and various electronic/spintronic device applications.

\section*{Acknowledgements}
The authors thank N. Li and D. L. Deng for valuable discussions. This work was supported by the MOST Project of China (Nos. 2014CB920903, 2016YFA0300603), the National Natural Science Foundation of China (Grant Nos. 11574029, and 11225418), and Singapore MOE Academic Research Fund Tier 1 (SUTD-T1-2015004).


%\balance


\footnotesize{
%\bibliography{rsc} %your .bib file

\begin{thebibliography}{10}
\expandafter\ifx\csname url\endcsname\relax
  \def\url#1{{\tt #1}}\fi
\expandafter\ifx\csname urlprefix\endcsname\relax\def\urlprefix{URL }\fi
\providecommand{\eprint}[2][]{\url{#2}}

\bibitem{1}
K.S. Novoselov, A. K. Geim, S. V. Morozov, D. Jiang, Y. Zhang,S. V. Dubonos,I. V. Grigorieva and A. A. Firsov, {\em Science\/}, 2004, {\bf 306}, 666--669


\bibitem{4}
M. Xu, T. Liang, M. Shi and H. Chen, {\em Chem. Rev.\/}, 2013, {\bf 113(5)}, 3766-3798


\bibitem{5} S. Z. Butler, S. M. Hollen, L. Cao, Y. Cui, J. A. Gupta, H. R. Guti\'errez, T. F. Heinz, S. S. Hong, J. Huang, A. F. Ismach, E. Johnston-Halperin, M. Kuno, V. V. Plashnitsa, R. D. Robinson, R. S. Ruoff, S. Salahuddin, J. Shan, L. Shi, M. G. Spencer, M. Terrones, W. Windl and J. E. Goldberger, {\em ACS Nano\/}, 2013, {\bf 7(4)}, 2898-2926


\bibitem{6} K. J. Koski and Y. Cui,  {\em ACS Nano\/}, 2013, {\bf 7(5)}, 3739-3743

\bibitem{3} A. K. Geim and K. S. Novoselov,  {\em Nat. Mater.\/}, 2007, {\bf 6(3)}, 183-191


\bibitem{57} D. K. Efetov  and P. Kim , {\em Phys. Rev. Lett.\/}, 2010, {\bf 105(25))}, 256805


\bibitem{58} J. Ye, M. F. Craciun, M. Koshino, S. Russo, S. Inoue, H. Yuan, H. Shimotani, A. F. Morpurgo and Y. Iwasa Y, {\em Proc. Natl. Acad. Sci.\/}, 2011, {\bf 108(32)}, 13002-13006


\bibitem{wolf} S. A. Wolf, D. D. Awschalom, R. A. Buhrman, J. M. Daughton, S. von Moln\'ar, M. L. Roukes, A. Y. Chtchelkanova  and D. M. Treger, {\em Science\/}, 2001, {\bf 294(5546)}, 1488-1495



\bibitem{hanw} W, Han, {\em APL Mater.\/}, 2016, {\bf 4}, 032401


\bibitem{9} O. Yazyev, {\em Phys. Rev. Lett.\/}, 2008, {\bf 101(3)}, 037203


\bibitem{h2} J. Zhou, Q. Wang, Q. Sun, X. S. Chen, Y. Kawazoe and P, Jena , {\em Nano Lett.\/}, 2009, {\bf 9(11)}, 3867-3870


\bibitem{caoc} C. Cao, M. Wu, J. Jiang and H. Cheng,  {\em Phys. Rev. B: Condens. Matter Mater. Phys.\/}, 2010, {\bf 81(20)}, 205424


\bibitem{h3} D. Soriano, N. Leconte, P. Ordej\'on, J. C. Charlier, J. J. Palacios and S. Roche, {\em Phys. Rev. Lett.\/}, 2011, {\bf 107(1)}, 016602


\bibitem{yccheng} Y. C. Cheng, Z. Y. Zhu, W. B. Mi, Z. B. Guo and U. Schwingenschloegl, {\em Phys. Rev. B: Condens. Matter Mater. Phys.\/}, 2013, {\bf 87(10)}, 100401


\bibitem{yjia} J. Du, C. Xia, W. Xiong, X. Zhao, T. Wang and Y. Jia, {\em Phys. Chem. Chem. Phys.\/}, 2016, {\bf 18}, 22678-22686



\bibitem{n4} F. Wu, C. X. Huang, H. P. Wu, C. H. Lee, K. M. Deng, E. J. Kan and P. Jena, {\em Nano Lett.\/}, 2015, {\bf 15(12)}, 8277-8281


\bibitem{sswang} Y. Wang, S. S. Wang, Y. Lu, J. Jiang and S. A. Yang, {\em Nano Lett.\/}, 2016, {\bf 16(7)}, 4576-4582


\bibitem{38} T. Cao, Z. Li and S. G. Louie, {\em Phys. Rev. Lett.\/}, 2015, {\bf 114(23)}, 236602


\bibitem{wyao} X. Wu, X. Dai, H. Yu, H. Fan, J. Hu and W. Yao, 2014 {\em arXiv--preprint:1409.4733\/}


\bibitem{40} J. Mahmood, E. K. Lee, M. Jung, D. Shin, I. Jeon, S. Jung, H. Choi, J. Seo, S. Bae, S. Sohn, N. Park, J. H. Oh, H. Shin and J. Baek,  {\em Nat. Commun.\/}, 2015, {\bf 6}, 6486.



\bibitem{44} P. E. Bl\"ochl, {\em Phys. Rev. B: Condens. Matter Mater. Phys.\/}, 1994, {\bf 50(24)}, 17953


\bibitem{45} G. Kresse and J. Hafner, {\em Phys. Rev. B: Condens. Matter Mater. Phys.\/}, 1993, {\bf 47(1)}, 558


\bibitem{46} G. Kresse and J. Furthm\"uller, {\em Comput. Mater. Sci.\/}, 1996, {\bf 6(1)}, 15-50


\bibitem{pbe} J. P. Perdew, K. Burke and M. Ernzerhof, {\em Phys. Rev. Lett.\/}, 1996, {\bf 77(18)}, 3865


\bibitem{hse} J. Heyd, G. E. Scuseria and M. Ernzerhof,  {\em J. Chem. Phys.\/}, 2003, {\bf 118(18)}, 8207-8215


\bibitem{41} R. Q. Zhang, B. Li and J. L. Yang, {\em Nanoscale\/}, 2015, {\bf 7(33)}, 14062-14070


\bibitem{43} S. Guan, Y. C. Cheng, C. Liu, J. F. Han, Y. H. Lu, S. A. Yang and Y. G. Yao, {\em Appl. Phys. Lett.\/}, 2015, {\bf 107(23)}, 231904

\bibitem{Liang2016} Z. H. Liang, B. Xu, H. Xiang, Y. D. Xia, J. Yin and Z. G. Liu, RSC Adv., 2016, 6, 54027

\bibitem{lzhu} T. Zhang and L. Zhu, {\em Phys. Chem. Chem. Phys.\/}, 2016


\bibitem{42} R. Q. Zhang, B. Li and J. L. Yang,  2015 {\em arXiv--preprint:1505.02584\/}


\bibitem{54} E. C. Stoner, {\em Proc. R. Soc. A\/}, 1938, {\bf 165}, 372-414 ; {\em Proc. R. Soc. A\/}, 1939, {\bf 169}, 339-371


\bibitem{hpan} H. Pan, J. B. Yi, L. Shen, R. Q. Wu, J. H. Yang, J. Y. Lin, Y. P. Feng, J. Ding, L. H. Van and J. H. Yin, {\em Phys. Rev. Lett.\/}, 2007, {\bf 99(12)}, 127201


\bibitem{56} H. W. Peng, H. J. Xiang, S. H. Wei,S. S. Li, J. B. Xia and J. B. Li, {\em Phys. Rev. Lett.\/}, 2009, {\bf 102(1)}, 127201


\bibitem{mak} K. F. Mak, K. He, C. Lee, G. H. Lee, J. Hone, T. F. Heinz and J. Shan, {\em Nat. Mater.\/}, 2013, {\bf 12}, 207


\bibitem{yzhang} Y. J. Zhang, T. Oka, R. Suzuki, J. T. Ye and Y. Iwasa, {\em Science\/}, 2014, {\bf 344}, 725


\bibitem{strain1} C. Lee, X. Wei, J. W. Kysar and J. Hone, {\em Science\/},2008, {\bf 321(5887)}, 385-388


\bibitem{strain2} K. S. Kim, Y. Zhao, H. Jang, S. Y. Lee, J. M. Kim, K. S. Kim, J. Ahn, P. Kim, J. Choi and B. H. Hong, {\em Nature\/}, 2009, {\bf 457(7230)} 706-710


\bibitem{strain3} S. Bertolazzi, J.  Brivio and A. Kis, {\em ACS Nano\/}, 2011, {\bf 5(12)}, 9703-9709


\bibitem{strain4} X. Peng, Q. Wei and A. Copple, {\em Phys. Rev. B: Condens. Matter Mater. Phys.\/}, 2014, {\bf 90(8)}, 085402


\bibitem{akin} D. Akinwande, N. Petrone and J. Hone, {\em Nat. Commun.\/}, 2014, {\bf 5}, 5678


\bibitem{chang} H. Y. Chang, S. Yang, J. Lee, L. Tao, W. S. Hwang, D. Jena, N. Lu and D. Akinwande, {\em ACS Nano\/}, 2013, {\bf 7(6)}, 5446-5452


\bibitem{yhui} Y. Y. Hui, X. Liu, W. Jie, N. Y. Chan, J. Hao, Y. T. Hsu, L. J. Li, W. Guo and S. P. Lau, {\em ACS Nano\/}, 2013, {\bf 7(8)},7126-7131

\bibitem{vdwh} A. K. Geim and I. V. Grigorieva, {\em Nature\/}, 2013, {\bf 499}, 419-425


\end{thebibliography}
%\bibliographystyle{rsc} %the RSC's .bst file

}

\end{document}